\documentclass[twocolumn,nobibnotes,aps,prx,floatfix,preprint]{revtex4-1}
\usepackage{mathptmx,graphicx}
\usepackage[utf8]{inputenc}
\usepackage{dcolumn}
\usepackage{braket}
\usepackage[T1]{fontenc}
\usepackage{float}
\usepackage{amsmath,amsfonts,amssymb,eqnarray}
\usepackage[dvipsnames]{color}
\usepackage[normalem]{ulem}
\usepackage{soul}
\usepackage{graphicx}
\usepackage{siunitx}
\usepackage{array,booktabs}
\usepackage[pdftex,hypertexnames=false,hidelinks]{hyperref}

\begin{document}
\author{M. Colautti$^{*1,2}$}
\author{F. S. Piccioli$^{*2}$}
\author{P. Lombardi$^{1,2}$}
\author{C. Toninelli$^{1,2}$}
\affiliation{$^1$\small European Laboratory for Non-Linear Spectroscopy (LENS), Via Nello Carrara 1, Sesto F.no 50019, Italy}
\affiliation{$^2$ \small National Institute of Optics (CNR-INO), Largo Enrico Fermi 6, Firenze 50125, Italy}
\affiliation{$^{\ast}$ \footnotesize These authors contributed equally to this manuscript}
\author{Z. Ristanovic$^{*3}$}
\author{A. Moradi$^{3}$}
\author{S. Adhikari$^{3}$}
\author{M. Orrit$^{3}$}
\affiliation{$^3$ \small Huygens-Kamerlingh Onnes Laboratory, LION, Postbus 9504, 2300 RA Leiden,
The Netherlands}
\author{I. Deperasinska$^{4}$}
\author{B. Kozankiewicz$^{4}$}
\affiliation{$^4$ \small Institute of Physics, Polish Academy of Sciences, Al. Lotnikow 32/46, 02-668 Warsaw, Poland}

\title{Laser-induced frequency tuning of Fourier-limited \\single-molecule emitters}

\begin{abstract}
The local interaction of charges and light in organic solids is the basis of distinct and fundamental effects. We here observe, at the single molecule scale, how a focused laser beam can locally shift by hundreds-time their natural linewidth and in a persistent way the transition frequency of organic chromophores, cooled at liquid helium temperatures in different host matrices. Supported by quantum chemistry calculations, the results are interpreted as effects of a photo-ionization cascade, leading to a stable electric field, which Stark-shifts the molecular electronic levels. The experimental method is then applied to a common challenge in quantum photonics, i.e. the independent tuning and synchronization of close-by quantum emitters, which is desirable for multi-photon experiments. Five molecules that are spatially separated by about 50 microns and originally 20 GHz apart are brought into resonance within twice their linewidth. Combining this ability with an emission linewidth that is only limited by the spontaneous decay, the system enables fabrication-free, independent tuning of multiple molecules integrated on the same photonic chip.
\end{abstract}
\maketitle
Single fluorescent molecules of polycyclic aromatic hydrocarbons (PAHs) embedded in crystalline organic matrices are widely considered as highly coherent, stable, and bright two-level quantum systems in the solid state \cite{Hettich2002a,Pototschnig2011,Wang2019, Orrit1990,Basche1992}. Therefore, they can be operated as single-photon sources\cite{Lounis2000}, combining high count rate and Fourier-limited linewidths \cite{Nicolet2007,Lettow2010a, Rezai2018}, or as nano-probes with exquisite sensitivity to electric fields, pressure, and strain \cite{Troiani2019, Tian2014}. PAH-based quantum devices can also be readily integrated in photonic chips \cite{Lombardi2017,Turschmann2017,Rattenbacher2019,Colautti2019}. A major advantage of PAH systems resides in the possibility to mass-produce nominally identical fluorescent molecules at low costs, and still obtain outstanding optical and opto-electronic properties. PAHs are indeed under investigation for their use also in organic solar cells \cite{Aumaitre2019}, as well as in super-resolution microscopy methods \cite{Yang2015}. Furthermore, Stark effect is a well established technique for tuning the molecular transitions\cite{Wild1992, Orrit1992,Tamarat1995}. Broad tunability has been demonstrated for instance in the case of anthracene nanocrystals (Ac NCXs) doped with dibenzoterrylene (DBT) molecules, exhibiting a quadratic Stark shift upon the application of a voltage bias (tuning range >\SI{400}{\giga\hertz}) \cite{Schaedler2019}. Also a large linear and homogeneous Stark shift has been achieved, embedding DBT in 2,3-dibromonaphtalene (DBN), suggesting the system as a promising nano-probe of electric fields and charges \cite{Moradi2019a}.
\par Such previous results demonstrate the potential of molecular-based techniques and, in principle, to go beyond single-emitter devices. However, current tuning methods should be scaled up and allow for local control of the transition frequency. This is crucial to optimize coupling of multiple emitters to photonic structures, as well as quantum interference among distinct sources on chip. To this purpose, electrical tuning of the emission wavelength for multiple sources is challenging to implement. Optical control might instead enable simple, fast, and independent frequency tuning of separated emitters. Recent promising works have shown laser-induced frequency tuning of epitaxial quantum dots' transitions. Physical manipulation of the sample is in this case confined within the laser beam diameter and localized tuning of the emitters is achieved via thermal annealing \cite{Fiset-Cyr2018}, mechanical expansion through phase-change materials \cite{Takahashi2013} or controlled strain through the host/substrate medium \cite{Grim2019}. Nevertheless, the observed decoherence and spectral diffusion of the emitters' linewidth can be detrimental in many applications.
\begin{figure*}
\centering
\includegraphics[width=0.85\textwidth]{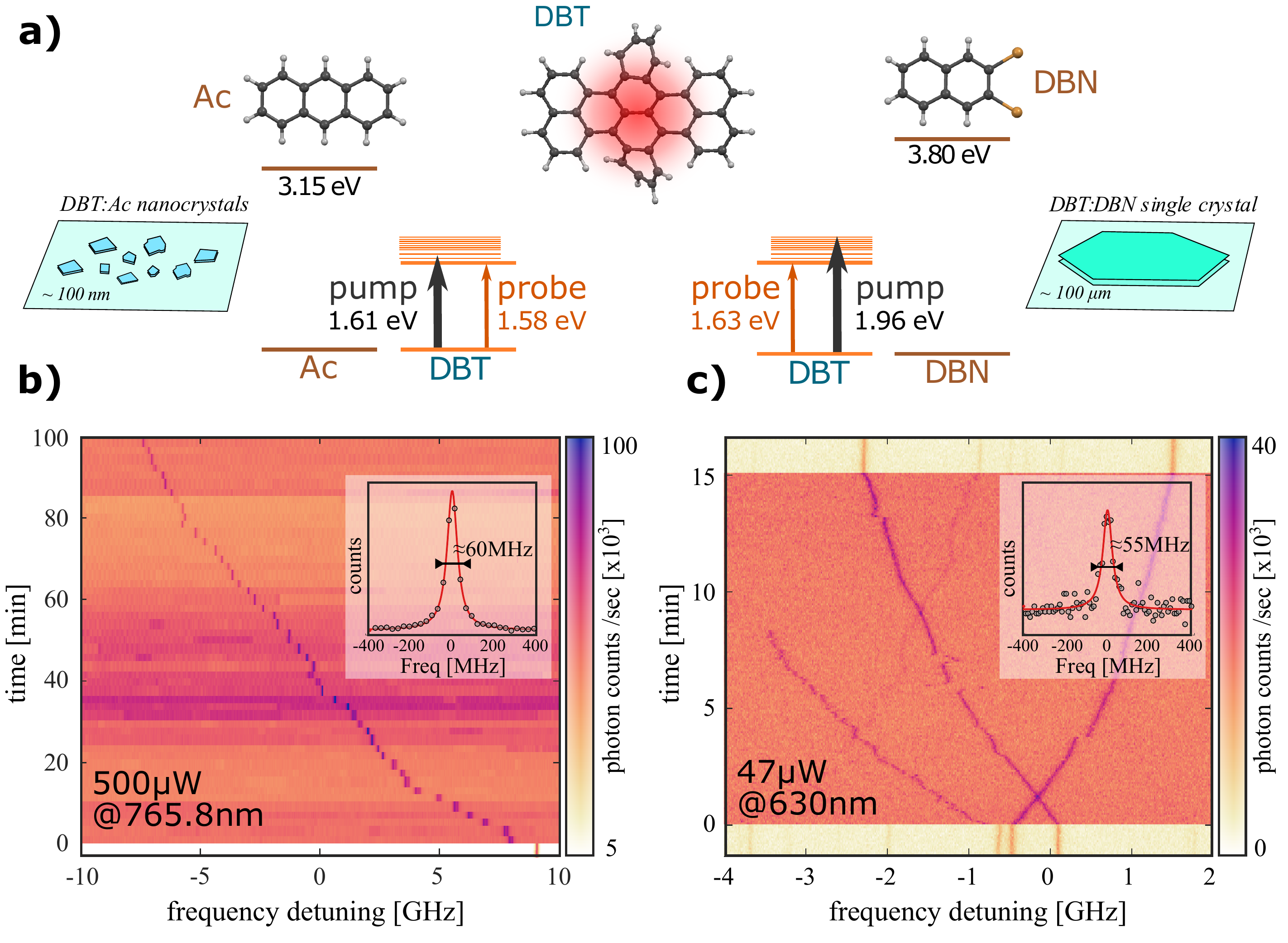}
\caption{a) Chemical structures of the guest (DBT) and host molecules (Ac, DBN) used in this study. The energies of the pump and probe beams used in the experiments are indicated for comparison. 
b,c) Laser-induced frequency shift for DBT in Ac (b) and DBT in DBN (c). The fluorescence count rate is plotted in color scale as a function of the excitation laser frequency, scanning over the molecular ZPL. The measurements are repeated in time, while the pump laser is switched on. The pump presence is recognized by the onset of a strong background signal, i.e., the light-yellow background represents scans without the pump beam. Insets in the panels (b) and (c) represent horizontal cuts and include the Lorentzian fits to the data. 
}
\label{fig:fig1}
\end{figure*}
\par In this work we demonstrate fabrication-free, micron-resolved, optical frequency tuning of individual DBT molecules. We attribute the laser-induced frequency tuning of the molecular Zero-Phonon Line (ZPL) to a local Stark shift, associated to optically-induced long-lived charge-separated states, which notably persists for at least several hours after the pump laser is switched off. We first demonstrate the versatility of the concept, studying DBT in two different types of molecular crystals. In Sec.~\ref{sec:characterization} we present a full characterization of the laser-induced frequency tuning. In particular, we show how the magnitude of the induced spectral shift can be extended over a broad frequency range, by acting on the pump laser parameters. We also verify that close-to-Fourier-limited emission and photo-stability are preserved at cryogenic temperatures after the tuning process. In Sec.~\ref{sec:spatial res.} we demonstrate the scalability of the presented method as we show independent tuning of individual emitters down to \SI{15}{\micro\metre} in proximity. In particular, we show ZPL frequency matching (within two linewidths) of five molecular emitters, all within a 50 $\times$ \SI{50}{\square\micro\metre} area and initially >\SI{20}{\giga\hertz} apart in frequency. Finally, in Sec.~\ref{sec:discussion} we propose a model based on the photo-ionization mechanism of charge generation and support it with low-level quantum chemistry calculations.
\par Overall, the possibility of using an all-optical approach to shift the transition frequency of individual emitters while maintaining coherent spectral properties offers a promising tool for applications in quantum nanophotonics. In particular, the presented results will benefit the realization of scalable quantum photonic devices and protocols based on multiple photons and multiple single-molecule emitters. More generally, the concept of laser-induced charge-separation via photoionization of molecules can have broader scope in applications of single-molecule emitters, for example in nanoscale sensing of charge-carrier generation and in understanding the photoconductivity and photorefraction of organic semiconductors.
\section{Laser-induced tuning of single-molecule emission}
The optically-induced frequency tuning of light emission is studied via single-molecule measurements on dibenzoterrylene (DBT) chromophores embedded in two different host systems, namely anthracene (Ac) nanocrystals doped with low DBT concentration, and 2,3-dibromonaphthalene (DBN) crystal flakes, doped with high DBT concentration. The molecular components of the two studied systems and their electronic energy levels are summarized in Fig.\thinspace{\ref{fig:fig1}}a. We also report additional results on a third system of DBT in naphthalene in the SI, showing that the effect applies to a broader range of organic crystals. The studied systems share the same guest molecule (DBT), but are known to have different response to the electric field. In particular, DBT:DBN exhibits a large linear Stark coefficient due to the broken inversion symmetry of guest DBT molecules \cite{Moradi2019a} while DBT:Ac shows a quadratic Stark coefficient \cite{Nicolet2007,Schaedler2019}. 
\par 
All the experiments are performed at liquid helium temperatures.
In these conditions, the purely electronic ZPL is about \SIrange{40}{60}{\mega\hertz} wide. The spectral line shape can be traced with a standard epifluorescence microscope by scanning the frequency of the CW excitation laser (named ``probe'' hereafter) around the resonant wavelength. The Stokes-shifted fluorescence, spectrally selected through a longpass filter, is correspondingly measured with an avalanche photodiode, giving access to the excited state occupation probability as a function of the excitation wavelength.
\par Optical frequency tuning of the narrow-band ZPL is achieved by focusing on the target crystalline regions a second CW laser (named ``pump'' hereafter), generally operating at a different wavelength and higher intensity.
In Fig.\thinspace{\ref{fig:fig1}}(b,c), real-time information on the laser-induced shift is provided by colour maps showing, as a function of time, the excitation spectra of DBT molecules, exposed to the pump laser for a long time interval. In Fig.\thinspace{\ref{fig:fig1}}a, simplified Jablonski diagrams are outlined, showing as well the energies of the pump and probe lasers  employed in the experiments. The probe laser is scanned over the ZPL transitions around \SI{785}{\nano\metre} for DBT:Ac (Fig.\thinspace{\ref{fig:fig1}}b) and \SI{756}{\nano\metre} for DBT:DBN (Fig.\thinspace{\ref{fig:fig1}}c), whereas the pump laser is centered at \SI{765.8}{\nano\metre} and \SI{630}{\nano\metre}, respectively. 
\par Zero-phonon lines corresponding to different molecules can thus be initially identified, followed during the burst duration above the background fluorescence, and recovered at different frequency positions after the exposure (Fig. 1c). It is worth noticing that after the pump laser illumination, DBT molecules preserve excellent spectral stability and close-to-lifetime limited resonance linewith. In particular, the spectral width amounts to $\sim\SI{60}{\mega\hertz}$ and $\sim\SI{55}{\mega\hertz}$ for DBT:Ac and DBT:DBN, respectively (see insets in Fig.\thinspace{\ref{fig:fig1}}). In the first case, the line broadening, compared to before the shift procedure, is unchanged and is attributed to the higher temperature of about \SI{3.5}{\kelvin} at which the experiment is held. Spectral wandering is present in these samples and is related to the fluctuations of the local electric field, due to charge migrations.
The energy shift in DBT emission persists even after the pump is turned off for about 24 h, ruling out the hypothesis of temporary heating effects (see Sec.~\ref{sec:discussion}). Furthermore, the frequency increment after each probe scan (single row in the colour map), of the order of few linewidths, hints at the possibility of a fine calibration of the effect, which is analysed in more details in Sec.~\ref{sec:characterization} and \ref{sec:spatial res.}. 
\par 
The number of DBT molecules exposed to the laser confocal illumination (beam diameter $d\sim\SI{1}{\micro\metre}$) is different in the two host-guest systems. On a sample of distributed DBT:Ac NCX (Fig.\thinspace{\ref{fig:fig1}}b), a single nanocrystal can be isolated within the beam area, which implies that typically only one molecule is resonant within a {\SI{500}{\giga\hertz} scan of the probe. However this cannot exclude the presence of broader molecules on the surface that could be weakly resonant with the intense pump laser}. The reported frequency shift is always towards larger wavelengths, in agreement with the characteristic quadratic Stark effect of DBT in Ac \cite{Schaedler2019}. On the other hand, within the highly doped DBN flakes, the confocal volume typically contains several chromophores within a \SI{10}{\giga\hertz} scan of the probe-laser frequency. As reported in Fig.\thinspace{\ref{fig:fig1}}c, shifts with opposite signs are observed within the sample of DBT:DBN, for which a large linear Stark coefficient has been reported \cite{Moradi2019a}.
Importantly, although each individual molecule may sense slightly different local electric fields (see the SI for more information), the fairly homogeneous response of most emitters in this well-defined system indicates that the laser-induced effect acts on all emitters within the diffraction limited-volume. These results jointly corroborate the interpretation of a laser-induced electric field buildup, leading to a Stark effect.
\par It is worth noticing that, when the pump is switched on, the measurements exhibit a strong background signal, which scales linearly with power. This light might be due to the presence of several DBT molecules within the illumination volume in the host matrices, weakly excited by the high-power pump laser. We note that such background shows much higher variability in the case of DBT:Ac than for DBT:DBN (Fig. 1(b,c), as expected, since fewer molecules are involved in the former system. The role of nearby non-resonant DBT molecules in the shifting mechanism is thoroughly discussed in Section \ref{sec:discussion} and in the Supplementary Information. Notably, the emission purity for single-photon-based applications is not compromised by such background levels (see SI, Fig.S3), since the pump power employed for single-photon source operation is much lower than that used to shift the ZPL frequency, yielding a negligible background level \cite{Lombardi2020}.
In the following, we discuss in details the shift dependence on pump-burst duration, power, and wavelength.

\section{Characterization of the light-induced frequency shift}\label{sec:characterization}
\begin{figure*}
\centering
\includegraphics[width=0.95\textwidth]{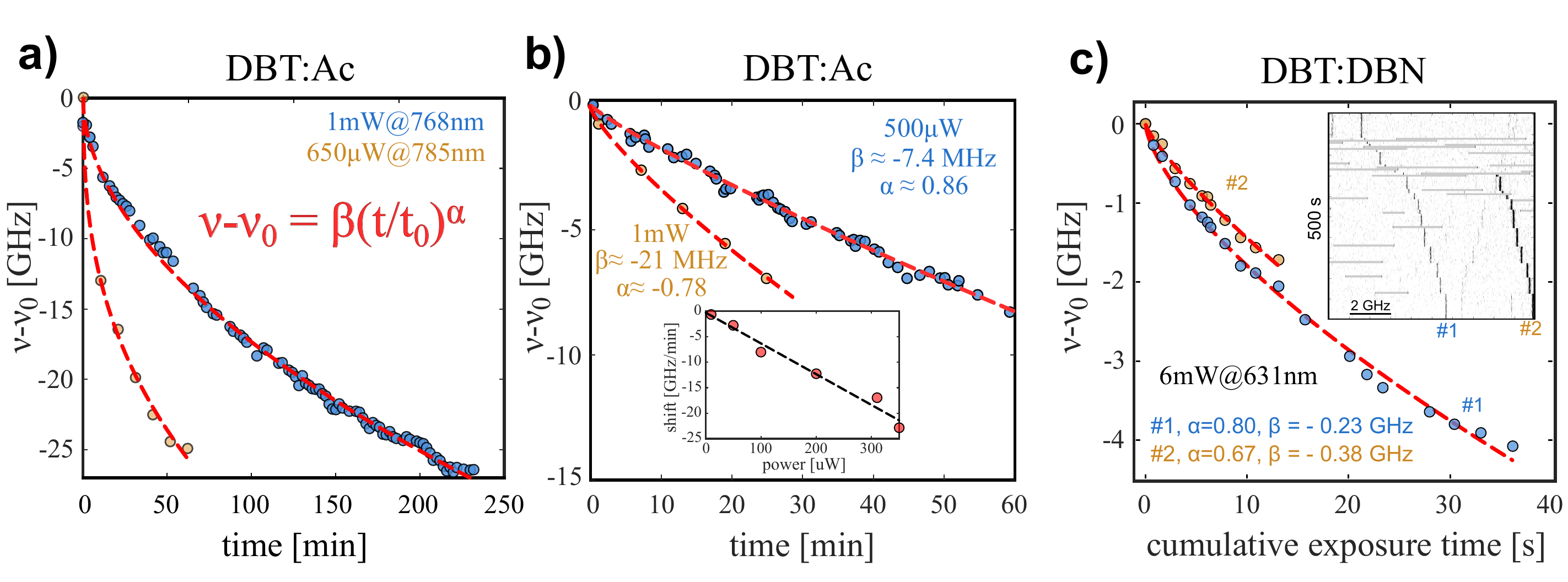}
\caption{Dependence of the laser-induced shift on the exposure time and pump power for DBT:Ac (a,b) and DBT:DBN (c). The data points are the ZPL resonant frequencies obtained from a Lorentzian fit to the real-time excitation spectra while the red dashed lines are power-law fits to the data. (a) spectral shifts for pump laser frequency around the 0-1 transition (blue points) and tuned 10 GHz below $\nu_0$ (yellow points) are compared. (b) Spectral shift dynamic for a DBT emitter pumped with different laser power. Inset: spectral shifts of a second molecule subject to several 2-min long bursts at increasing power show linear dependence on pump beam power. (c) Spectral shift as a function of cumulative exposure time for two DBT emitters (indicated as \#1 and \#2) in DBN. Inset: Frequency scans shown in real time (one linescan is \SI{5}{\second}). The gray horizontal lines indicate intermittent, short pump laser exposures (\SIrange{0.5}{3}{\second}).}
\label{fig:fig2}
\end{figure*}
The results on the dynamics of the optical tuning are analysed in Fig.\thinspace{\ref{fig:fig2}}, where the frequency shifts of individual molecules in the two host-guest systems are displayed (Fig. 2a,b for DBT:Ac and Fig. 2c for DBT:DBN). The experimental data represent the ZPL central frequency as a function of the cumulative pump exposure time, and are extracted from the temporal map of the probe excitation spectrum via Lorentzian fits. 
As the pictured data suggest, in each observed case the frequency shift slows down for long exposure times, with a characteristic time that typically varies among emitters and is, on average, different in the investigated matrix hosts.
In general, such behaviour can be fairly well described as a power-law function of time \cite{nicolet2013} (dashed red lines in the figure are fits to the experimental data):
\begin{equation}\label{eq:powerLaw}
    \nu(t)=\nu(0)+\beta(P)(t/t_0)^{\alpha}\,,
\end{equation}
where $\nu$ is the molecule ZPL central frequency, $\nu(0)$ its initial value, $t$ the cumulative pump-exposure time, $t_0 = \SI{1}{\second}$ the time normalization constant, $\alpha$ indicates the power exponent and the parameter $\beta(P)$ describes the extent of the frequency shift that can be achieved in a given exposure time. The specific dynamic parameters for the molecule's ZPL can vary among the two host matrices and even between different molecules within the same sample. This might be due to local differences in DBT concentration/orientation and to local crystalline impurities.

Interestingly, no clear difference is observed in the induced shifts upon variation of the pump laser wavelength over a range of several nm, (see also the SI). This suggests that the shift phenomenon does not directly involve photon absorption by the shifting molecule. Indeed this eventuality is ruled out by tuning the pump laser to the red of the molecule's ZPL, where no excited states exist. Fig.\thinspace{\ref{fig:fig2}}a shows in yellow dots the results of such experiment, performed interrupting the pump illumination at regular intervals for about one-hour time. Despite a pump detuning of \SI{10}{\giga\hertz}, a large frequency shift of >\SI{20}{\giga\hertz} is observed. 
Every DBT emitter conversely shows a clear dependence on the pump-laser power. Specifically, when comparing shifts induced in the very same molecule, the parameter $\beta$ is strictly increasing with the pump laser power and, for short enough illumination times, has a linear dependence. In Fig.\thinspace{\ref{fig:fig2}}b two frequency shifts operated on the same molecule, exposed to different values of the pump laser power, are compared. In particular, for the higher power the ZPL is probed during brief interruptions of the pump illumination every few minutes, since a real-time measurement is precluded by the increased background fluorescence.
In the inset of Fig.\thinspace{\ref{fig:fig2}}b, a collection of data is displayed for a molecule in a different nanocrystal, which exhibits considerably larger frequency shifts and enables a characterization for low pump power values. In this case the experimental data represent the overall frequency shifts observed after single illumination bursts of 2 minutes each with increasing pump power, illustrating the linear behaviour of $\beta(P)$ for short illumination time. This result rules out the possibility of the direct two-photon-induced photoionization of the matrix, which would have a quadratic power dependence.

Qualitatively similar trends are reported for DBT:DBN in Fig.\thinspace{\ref{fig:fig2}}c, where two DBT emitters in the same DBN crystal are simultaneously shifted using a sequence of short, high power, laser bursts. The burst duration varies between \SIrange{0.5}{3}{\second} with constant laser power of \SI{6}{\milli\watt}. While one of the molecules undergoes photobleaching or large spectral jump during the process, the other acumulates a total shift of \SI{4}{\giga\hertz} in less than \SI{40}{\second} of cumulative excitation. More results on intermolecular differences in DBT:DBN system are available in the SI.

\section{Tuning molecules into resonance}\label{sec:spatial res.}
\begin{figure}[hb]
\includegraphics[width=0.45\textwidth]{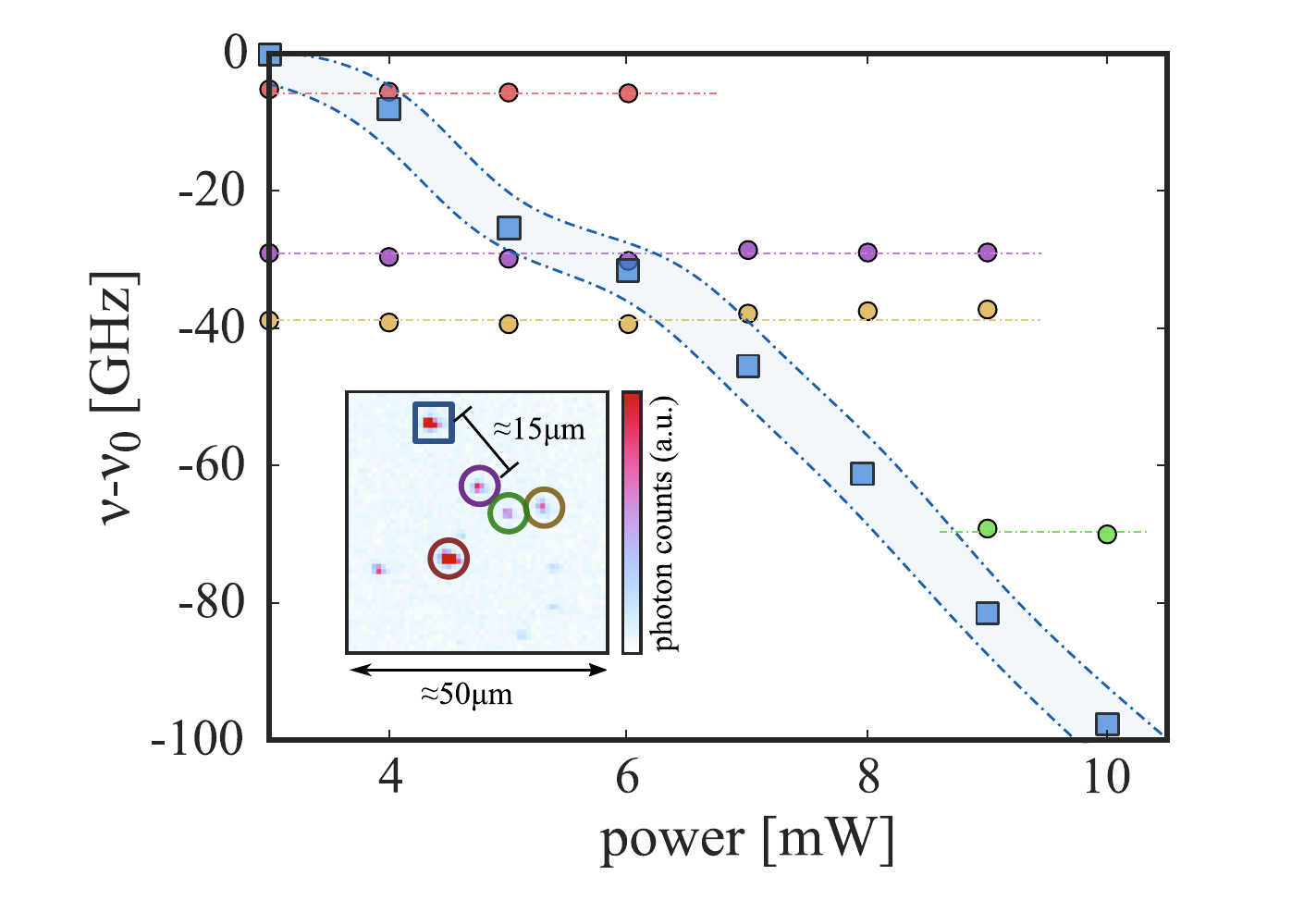}
\caption{Localized frequency tuning. The central frequency of the ZPL is reported for several molecules as a function of the pump laser power, which only affects one of them (blue squares). The spatial distribution of the doped nanocrystals is given in the inset fluorescence map, reveling that as-close-as-\SI{15}{\micro\metre} molecules are completely unaffected by the manipulation.
}
\label{fig:fig3}
\end{figure}
\begin{figure*}
\includegraphics[width=0.95\textwidth]{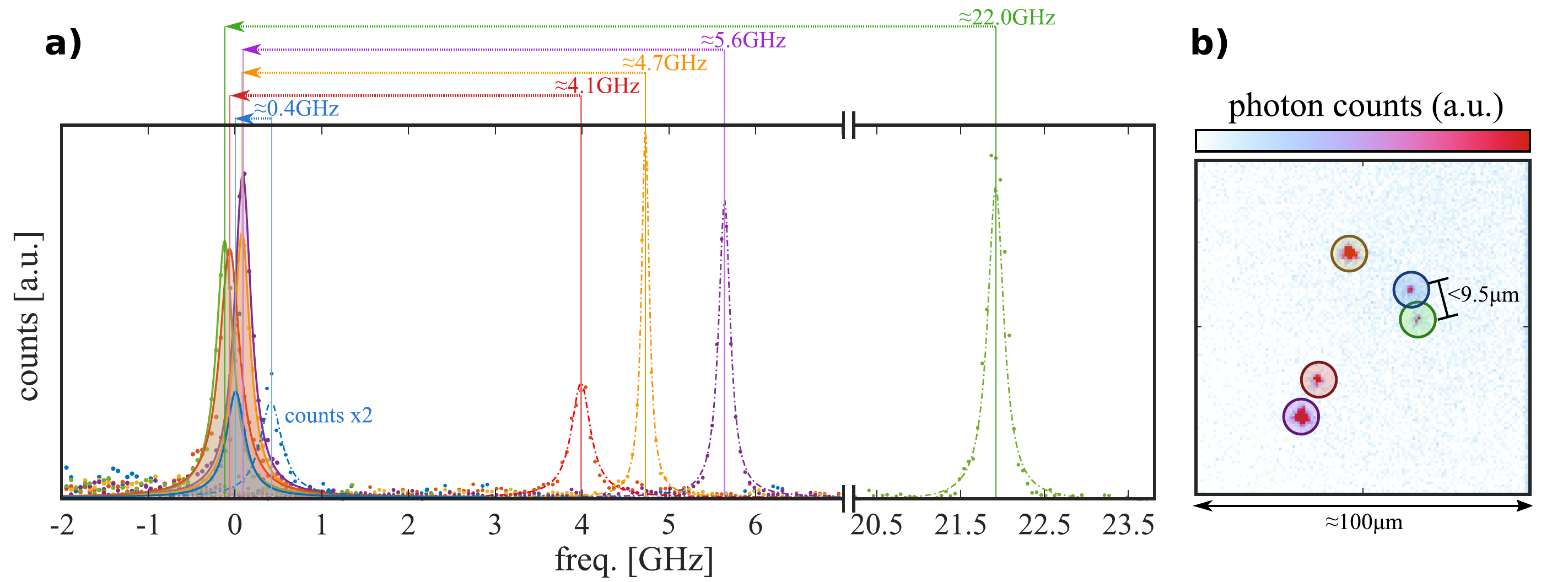}
\caption{Frequency matching of five distinct molecules can be observed in the excitation spectra displayed in a) as solid lines with underlying shaded areas. This exceptional condition is the result of a set of individual shifting applied to the original configuration, corresponding to the spectra reported as dashed dotted color lines. b) fluorescence map showing the simultaneous excitation of all nanocrystals when the probe is centered around $\nu_0$ and applied in wide-field configuration.}
\label{fig:fig4}
\end{figure*}
Remarkable advantages of the all-optical frequency tuning presented here are its spatial resolution and the flexibility of operation. Indeed, the effect is directly activated on the emitting sample with no need for additional fabrication and, being confined to the confocal beam volume, it enables spatial resolution at the micron level. On the other side, once doped with single-molecule concentration, homogeneously deposited on the substrate via dropcasting and dessication \cite{pazzagli2018}, or even integrated in nanophotonic structures\cite{Colautti2019}, DBT:Ac nanocrystals allow for spatial isolation of individual molecules which can thus be selectively tuned by laser-focusing. This is shown in Fig.\thinspace{\ref{fig:fig3}} where the ZPL peaks of five nearby molecules, each contained in a different nanocrystal (data of different color), are probed at regular time intervals while exposing only one of the molecules to 2 minutes long laser pulses with increasing power. The corresponding molecule (blue in the figure) undergoes a large shift of about \SI{100}{\giga\hertz} ($\sim\SI{0.20}{\nano\metre}$), whereas the transition frequency of others is not affected. In the inset, a wide-field fluorescence map shows the location of the respective nanocrystals, labeled with the same colour-code as in the main figure. Their relative distance amounts to approximately \SI{10}{\micro\metre} and is determined by the NCX suspension density before deposition. We note in passing that the limit to the spatial resolution of this approach has not been touched here and is likely determined by the specific charge migration mechanism in the sample (see Section \ref{sec:discussion}).
\par
Finally, we show the scalability of the proposed tuning approach, by shifting independently in a controlled way the ZPL of up to $5$ close-by molecules, which are all brought on resonance within twice their linewidth. As in the case of the measurements shown in Fig.\thinspace{\ref{fig:fig4}}a, the emitters are embedded in different nanocrystals. Moreover their emission linewidths, shown with the dashed lines in Fig.\thinspace{\ref{fig:fig4}}, are originally well-separated by \SI{20}{\giga\hertz}.
The nanocrystals are addressed by the confocal pump-beam spot one after the other, independently, in order to shift their initial transition frequencies to a common targeted value. The latter, because of the dominant quadratic Stark effect in the Ac-matrix, is necessarily chosen as $\nu_0\leq\nu_{i}$, where $i$ labels the molecules' original emission frequencies. The final excitation spectra are reported in Fig. \ref{fig:fig4} using solid lines with shaded underlying areas in different colors, demonstrating fine-matching of five molecules, within about a two-linewidth range. Total shifts of up to about \SI{20}{\giga\hertz} can be estimated by comparing the final values with the initial ones. In Fig.\thinspace{\ref{fig:fig4}}b, the relative wide-field fluorescence map is reported, showing the simultaneous excitation of all molecules by means of the probe beam centered at the targeted frequency $\nu_0$.
Correspondingly, the bright spots highlight the spatial configuration of the synchronized nanocrystals. We finally observe that similar results have been obtained on different substrates, with nanocrystals deposited either on gold (SI) or on sapphire (Fig.\thinspace{\ref{fig:fig4}}).
\par
The demonstated results clearly show together several advantages of the proposed tuning approach: scalability, flexibility in the choice of substrate, high spectral and spatial resolution. We believe such characteristics will be crucial for the coupling of multiple emitters together in linear optical computing or simulation experiments and for the integration of molecules into resonant photonic structures.

\section{Photoionization model of the laser-induced charge separation}\label{sec:discussion}
\begin{figure}[h!]
\includegraphics[width=0.45\textwidth]{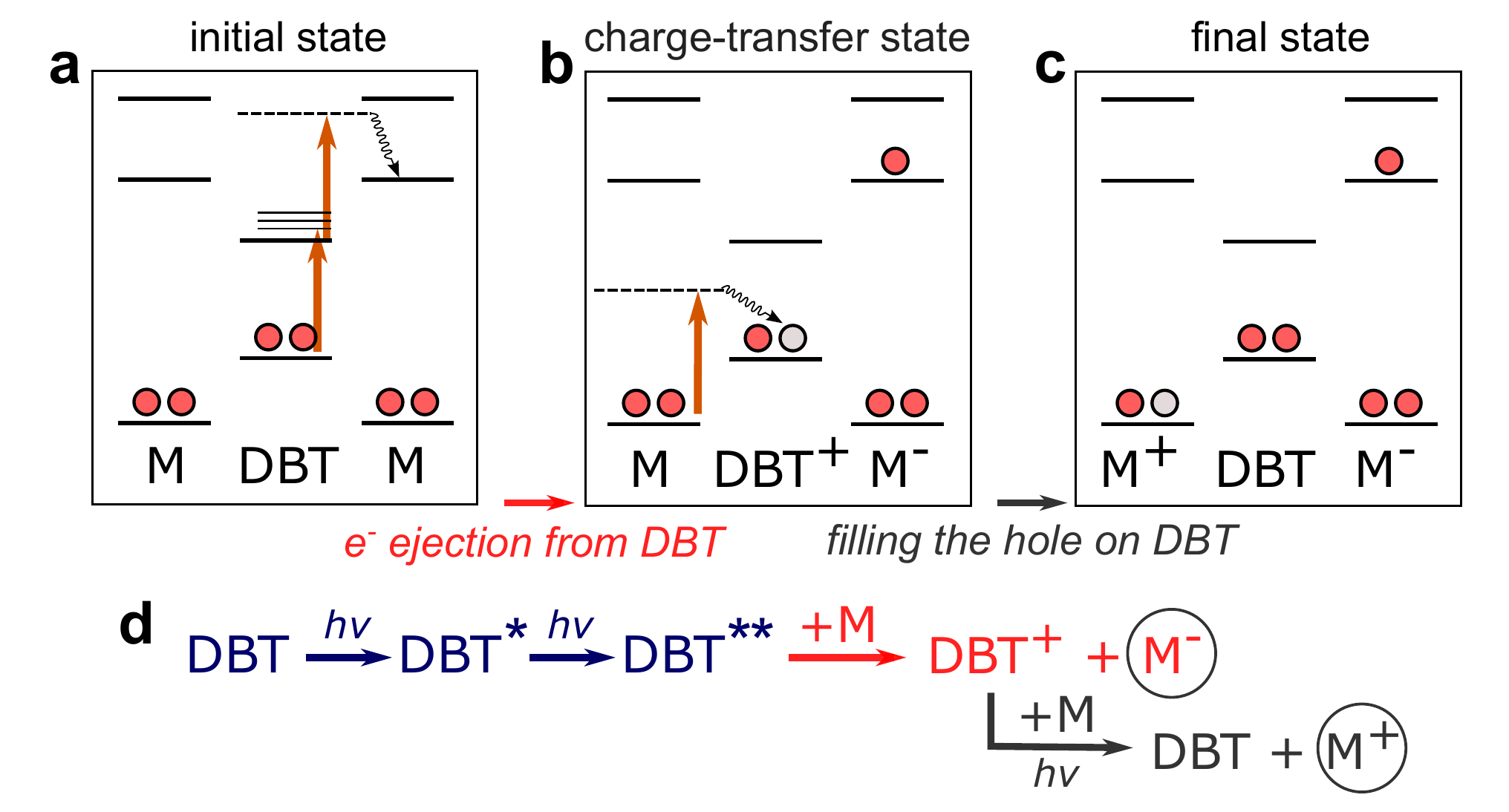}
\caption {The proposed mechanism of the laser-induced charge generation. a-c) Electronic energy diagrams describing a) the initial electronic state and local excitation of DBT (red arrows), followed by electron ejection to the matrix (wavy line), (b) charge-transfer state and filling the hole on DBT, and (c) resulting final state of the system. Transitions induced by a single photon are shown with red arrows, virtual states are indicated by broken lines, and non-radiative CT transitions by wavy lines. Red and gray circles designate electrons and holes, respectively. M = matrix molecule (DBN or Ac) d) Photoionization scheme that represents the path sketched in panels (a-c); $h\nu$ = energy of a pump-beam photon; excited electronic states are represented with the star symbol.
}
\label{fig:fig5}
\end{figure}

In this section we propose a mechanism for our hypothesis of the laser-induced charge separation through photoionization of DBT molecules. We postulate a cascade photoionization mechanism for the buildup of the local electric field, via formation of long-lived charge-separated states. The proposed model is based on the following assumptions: 
\begin{enumerate}
\item The laser-induced excitation of DBT molecules leads to their photoionization by electron ejection.
\item The process results in long-lived charge-separated states in the matrix. The initially excited DBT molecules may be regenerated to their neutral state.
\item The photoionization of DBT results in mobile charge carriers (electrons and holes, with different mobilities) and their transport to larger distances (in the order of tens of nanometers at least). 
\item The charge separation and trapping of electrons and holes result in the buildup of a static electric field which shifts DBT's ZPLs through Stark effect (quadratic for DBT:Ac and linear for DBT:DBN).
\end{enumerate}
The laser-induced charge separation mechanism is outlined in Fig.\thinspace{\ref{fig:fig5}}. The initial step of the charge generation process is the photoionization of DBT molecules by intense pump light (Fig.\thinspace{\ref{fig:Fig5}}a). This process requires at least two laser photons: a first DBT excitation with one pump photon (\SI{1.58}{\electronvolt} for DBT:Ac or \SI{1.96}{\electronvolt} for DBT:DBN) to its first excited electronic state (S1), followed by a further excitation with a second pump photon (red arrows in Fig.\thinspace{\ref{fig:fig5}}a). Such a highly excited electronic state of DBT can presumably eject an electron to the matrix (M) and give birth to a charge-separated state (DBT\textsuperscript{+}/M\textsuperscript{-}). 
Finally, the hole left on the DBT molecule may be transferred to the matrix through a single-photon excitation (Fig.\thinspace{\ref{fig:fig5}}b), yielding back a neutral DBT molecule and a matrix cation (M\textsuperscript{+}) (Fig.\thinspace{\ref{fig:fig5}}c). Figure \ref{fig:fig5}d summarizes the whole photoionization process.
\par
The electron and hole migration in the matrix is possibly further activated by the pump beam, leading to charge diffusion and to charge-separated states (M\textsuperscript{+} and M\textsuperscript{-}) far enough from each other so that the recombination probability becomes negligible. It is important to realize that, if the initial DBT molecule is regenerated to its neutral state, the process can be repeated many times. In effect, each DBT molecule may give rise to a macroscopic charge distribution in its environment. The resulting charge density buildup shifts ZPLs of DBT probe molecules via Stark effect. 

\par Our photoionization scheme is inspired by earlier models of photoconductivity in organic solids. The inclusion of larger PAH impurities into anthracene crystal, such as tetracene, significantly reduces the activation energy for the charge-carrier transport.
\cite{Silinsh1980, Karl1974, Hoesterey1963}.
Likewise, the proposed charge separation mechanism is the key to the photorefractive effect in multi-component photorefractive materials doped with optically active organic molecules \cite{Moerner1994_ACS}.
\par The suggested model is consistent with the continuous shift of the molecule's ZPL resonance, as well as with its narrow line, which rules out any heating-induced rearrangements. The Stark shift associated to a net field build-up well captures a shift that can have opposite signs for DBT:DBN, while is only towards longer wavelengths for DBT:Ac. Further details on the consistency between the model and the experimental observations are provided in the SI.
\par
Finally, simple quantum chemistry calculations have been performed, which qualitatively support our photoionization model. All three steps presented in Fig.\thinspace{\ref{fig:fig5}} are found to be energetically feasible according to time-dependent density functional theory (TDDFT/B3LYP). For the details of the calculations, the reader is referred to the SI. Beyond the first electronic excitation, several higher electronic states of DBT are found above \SI{3.0}{\electronvolt}, suggesting that excitation of DBT with two red photons is likely to happen  under experimental conditions of strong illumination. 
According to the calculations, the following step of electron transfer to the matrix could be phonon-assisted for DBT in Ac, while a direct excitation of the charge-transfer (CT) state may operate as well for the DBT:DBN system. 
The transport of charges is further supported by the presence of many host molecules in the crystals, which results in the formation of a quasi-band structure of CT-type electronic states (see the SI). \cite{Schwoerer2007}
Concerning the regeneration of neutral DBT molecules, several CT transitions are found from Ac to DBT\textsuperscript{+} within the energy range of our excitation (1.6 eV). 
Lastly, the further transfer of charge carriers through the matrix to distances larger than several unit cells is assumed to take place through interaction of charged species with red light.  Indeed, we find that Ac\textsuperscript{+}, Ac\textsuperscript{-} and DBT\textsuperscript{+} can all absorb red photons, which is in line with other theoretical results. \cite {Malloci2007}
\section*{Conclusions}
We have demonstrated optically-induced tuning of light emission from single DBT molecules in two different types of molecular crystals that are of interest for molecule-based integrated quantum devices. 
Tuning by up to \SI{0.2}{\nano\metre} has been observed, persisting for a macroscopic time (several hours), which allows the successive manipulation of several independent molecules. Because of the predictability of the frequency shift we were able to tune $5$ independent molecules on the same ZPL emission wavelength within twice their linewidth. Such fine control of an emitter's transition frequency, together with the achievable tuning range, might empower the presented technique in any context involving molecular emitters coupled to resonating photonic structures and/or multiple sources emitting at the same wavelength, such as in optical quantum computation, simulation and communication.
\par
We emphasize the potential applicability of our methodology to a broad range of host/guest molecular systems. Studying light-induced charge dynamics with single-molecule sensitivity could shed light on the microscopic mechanisms behind photoconductivity in organic solids and open new venues for the exploitation of single molecules in quantum technologies and molecular electronics. Many possibilities to further fine-tune and manipulate optical transitions of molecules can be envisaged in this context. Finally, tuning light emission of many single emitters by using this simple optical approach may become an important step in integrating single emitters into robust quantum protocols based on molecules.

\section*{Acknowledgements}
Nico Verhart is acknowledged for performing the initial experiments on DBT in naphthalene. C.T. wishes to thank P. Foggi and F.S. Cataliotti for useful discussions. The authors acknowledge the EraNET Cofund Initiative QuantERA under the European Union’s Horizon 2020 Research and Innovation Programme (ORQUID, grant agreement no. 731473). The Netherlands Organization for Scientific Research and NWO-Physics are acknowledged for funding for A.M., S.A., and Z.R. Theoretical calculations were performed at the Interdisciplinary Centre of Mathematical and Computer Modelling (ICM) of Warsaw University under the computational grant no. G-32-10.
\par
M.C., F.S.P., and Z.R. contributed equally to the manuscript. M.C., F.S.P., and P.L. performed the experiments on DBT in anthracene. Z.R., A.M., and S.A. performed the experiments on DBT in 2,3-dibromonaphthalene and naphthalene. I.D. performed the quantum chemistry calculations and together with B.K. contributed to the photoionization model. C.T. and M.O. coordinated the research. All authors contributed to the final manuscript.

%
\end{document}